\documentclass{webofc}
\usepackage[varg]{txfonts}   

\newcommand{\mnras}{MNRAS, }
\newcommand{\apj}{ApJ, }
\newcommand{\aap}{A\&A, }
\newcommand{\apjl}{ApJL, }

\newcommand{\be}{\begin{equation}}
\newcommand{\ee}{\end{equation}}

\newcommand{\apjs}{{\it ApJS, }}

\woctitle{Instabilities and structures in protoplanetary disks}
\begin{document}
\title{Instabilities at planetary gap edges in 3D self-gravitating
  disks}
%
%

\author{Min-Kai Lin\inst{1}\fnsep\thanks{\email{mklin924@cita.utoronto.ca}}
}

\institute{Canadian Institute for Theoretical Astrophysics 
  , 60 St. George Street,
  Toronto, Ontario, M5S 3H8,
  Canada  }

\abstract{Numerical simulations are presented to study the
  stability of gaps opened by giant planets in 3D
  self-gravitating disks. In weakly self-gravitating
  disks, a few vortices develop at the gap edge and merge on orbital 
  time-scales. The result is one large but weak vortex with Rossby number -0.01. 
  In moderately self-gravitating disks, more vortices develop and their
  merging is resisted on dynamical time-scales. Self-gravity can sustain
  multi-vortex configurations, with Rossby number -0.2 to -0.1, over a
  time-scale of order 100 orbits. Self-gravity also enhances the vortex
  vertical density stratification, even in disks with
  initial Toomre parameter of order 10. However, vortex formation
  is suppressed in strongly self-gravitating disks and replaced by a
  global spiral instability associated with the gap edge which
  develops during gap formation. 
}
\maketitle
\section{Introduction}\label{intro}
Gaps induced by planets in protoplanetary
disks can become dynamically unstable if the disk viscosity is
sufficiently small \citep{koller03}. 
This is because planetary gap edges are
associated with potential vorticity or vortensity extrema \citep{li05,lin10}, 
the existence of which is necessary for instability
\citep{lovelace99}. Gap edges may undergo vortex formation in weakly
self-gravitating disks (associated with vortensity minima)
or a spiral instability (associated with vortensity maxima) in strongly
self-gravitating yet Toomre-stable disks
\cite{lin11a,lin11b}. Development of such instabilities can 
significantly affect planetary migration
\cite{lin_pap_12} and dust evolution \cite{inaba06}.  
These studies have employed 2D disk models, but gap edges
have characteristic widths of the disk local scale-height, so it is necessary
to extend the study of gap stability to 3D. 

\section{Numerical simulations with \texttt{ZEUS-MP}}\label{sims}
The system is an inviscid, non-magnetized 3D fluid disk embedded with a giant
planet of mass $M_p$, both rotating about a central
star of mass $M_*$. Spherical co-ordinates $(r,\theta,\phi)$
centered about the star are adopted. Units are such that $G=M_*=1$
, where $G$ is the gravitational constant.  

The disk is governing by the Euler equations coupled with
self-gravity through the Poisson equation. The equation of state is
locally isothermal, so the sound-speed is $c_s=H\Omega_k$, where
$H=hR$ is the isothermal scale-height with constant aspect-ratio $h$, 
$\Omega_k^2 \equiv GM_*/R^3$ and $R=r\sin{\theta}$. Each disk model is
labelled by its minimum Keplerian Toomre parameter
$Q_0$, located at the outer disk boundary. The planet is regarded as
an external potential and held on a circular Keplerian orbit of
radius $r_p$ at the midplane. Time is quoted in units of $P_0\equiv
2\pi/\Omega_k(r_p)$. The Hill radius $r_h\equiv(M_p/3M_*)^{1/3}r_p$ is used in some of the plots.  

The self-gravitating hydrodynamic equations are evolved with the \texttt{ZEUS-MP} 
finite difference code \cite{hayes06}. The computational domain, unless
otherwise stated, is
$r\in[1,25],\,\theta\in[\theta_\mathrm{min},\pi/2]$ and
$\phi\in[0,2\pi]$ where $\tan{(\pi/2-\theta_\mathrm{min})}=2h$.
Boundary conditions are outflow in $r$, reflecting in $\theta$ and
periodic in $\phi$. The numerical resolution is $N_r\times
N_\theta\times N_\phi=256\times32\times512$. See \cite{lin12b} for
further details for the simulation setup.     

\section{Results}

\subsection{Weakly self-gravitating disks ($Q_0=8$)}\label{q8}
Two simulations with $Q_0=8$ were run, one with self-gravity and the
other without. $M_p=0.002M_*$ and $h=0.07$ are adopted. Both cases 
developed $2$---$3$ vortices early on but the quasi-steady state is
a single vortex with $Ro\sim -0.01$,  where the Rossby number is
defined as $Ro~ \equiv ~ \omega_z/\langle2\Omega\rangle$, where
$\omega_z$ is the absolute vertical vorticity and $\langle\Omega\rangle$ is the
azimuthally-averaged angular speed.    

\begin{figure*}
  \centering
  \includegraphics[scale=.4,clip=true,trim=0cm 0.25cm .0cm
    0.65cm]{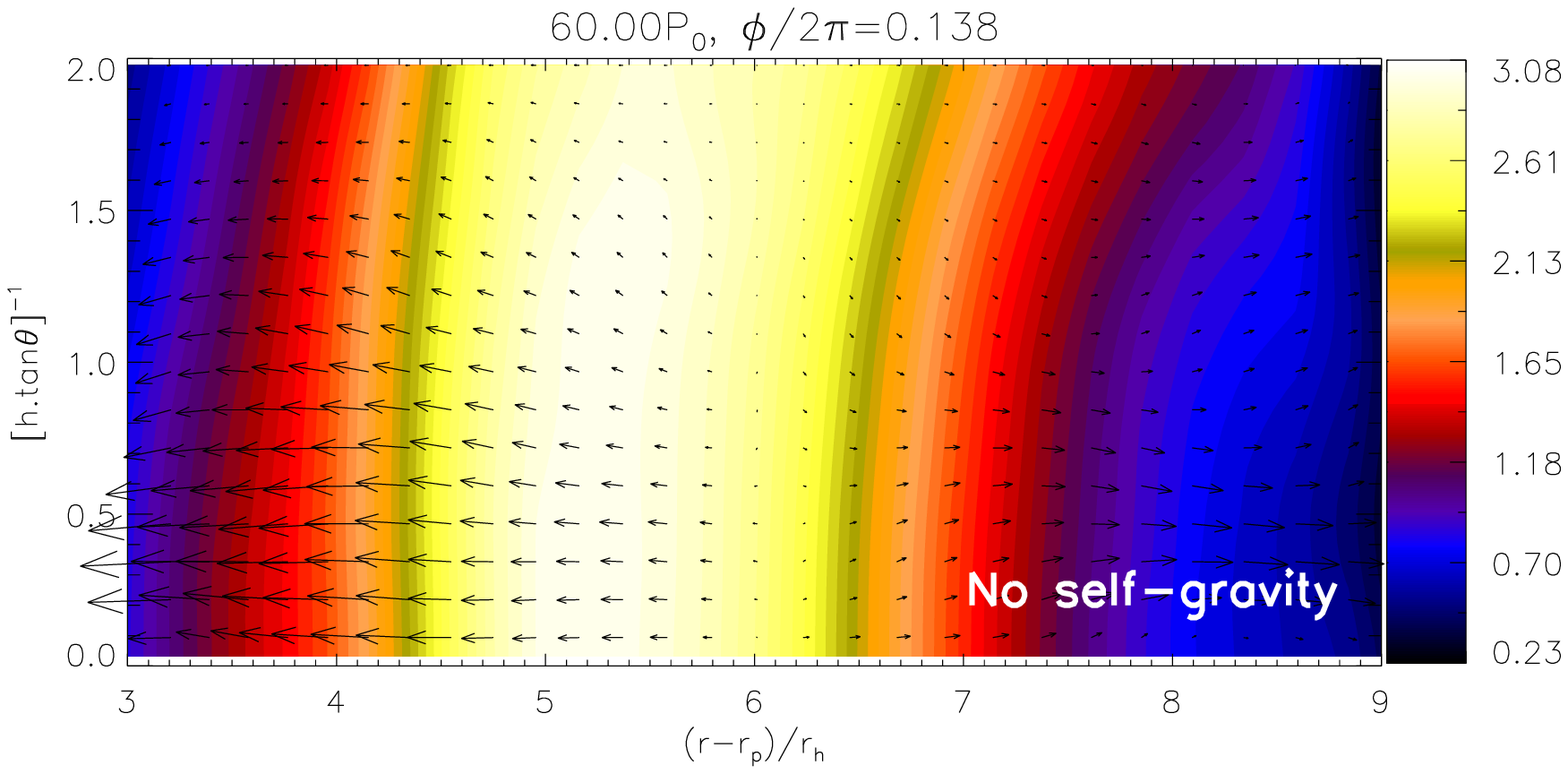}\includegraphics[scale=.4,clip=true,trim=1.45cm
    0.25cm .0cm
    0.65cm]{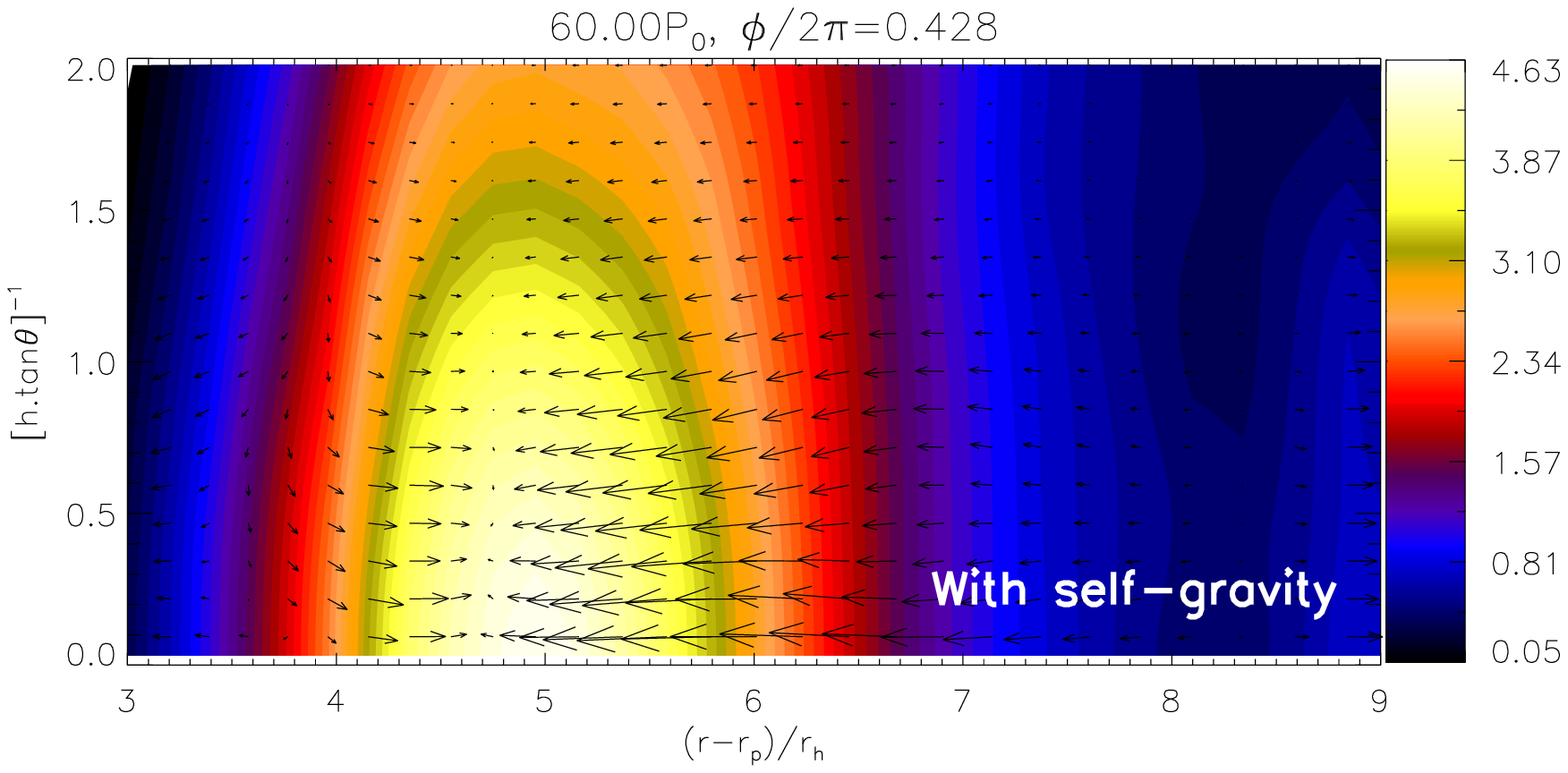}
  \caption{Relative density perturbation in the $(r,\theta)$ plane,
    chosen at the azimuth that intercepts the vortex centroid formed in
    two disk models with $Q_0=8$. The perturbed meridional flow is
    also shown. 
  }
  \label{vortex8}
\end{figure*}

The vortices differ noticeably in the $(r,\theta)$ plane. This  
is shown in Fig. \ref{vortex8}. Self-gravity enhances the vertical
density stratification of the vortex, with the midplane density
enhancement being $\simeq 50\%$ larger than that in the non-self-gravitating
run. The initial Keplerian Toomre parameter at the radius of vortex
formation is about $10$, but even this is sufficient to affect the
vortex vertical structure. 
The creation of vortensity minima lowers the Toomre parameter, which
further decreases with vortex formation since they are 
over-densities. Thus, self-gravity can become important in the
perturbed state even if it is negligible initially.   

\subsection{Moderately self-gravitating disks ($Q_0=3$)}\label{q3a}
\begin{figure*}[!ht]
  \centering
  \begin{minipage}{0.61\linewidth}
    \includegraphics[scale=.39,clip=true, trim = 0cm 0.0cm 0cm 0.92cm
    ]{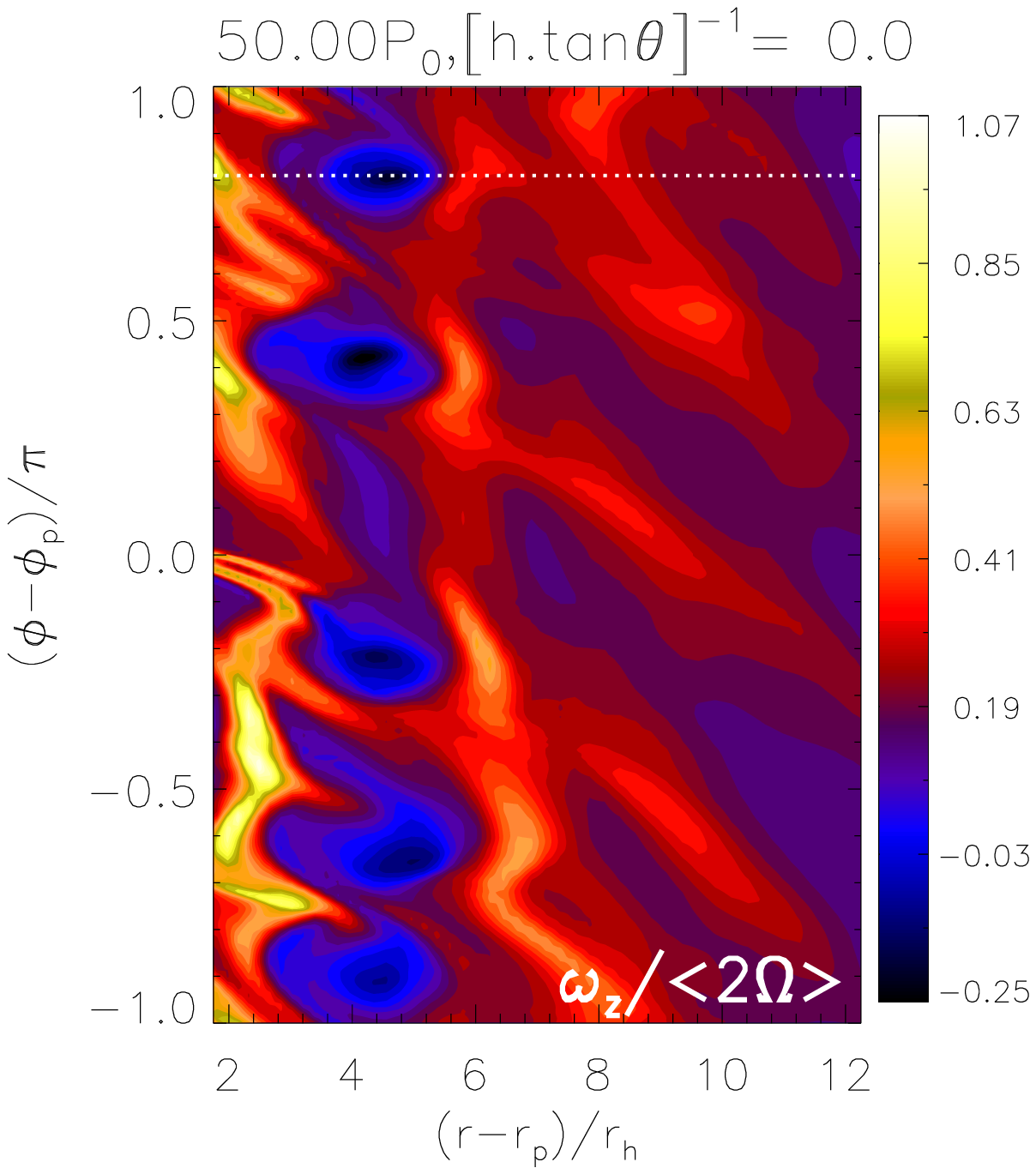}\includegraphics[scale=.39,clip=true, trim = 2.2cm 0.0cm 0cm 0.92cm
    ]{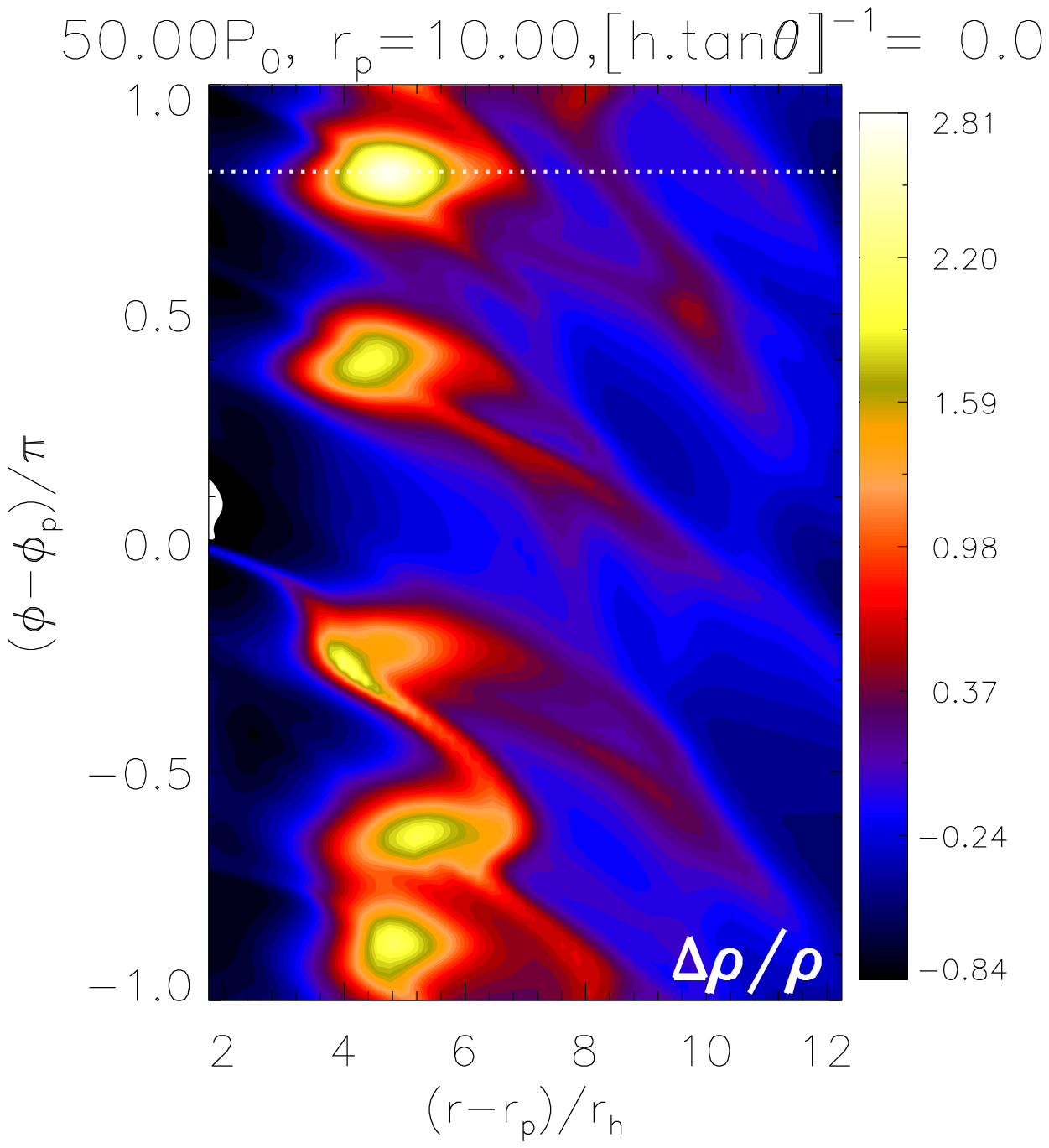} 
    \end{minipage}
  \begin{minipage}{0.38\linewidth}
    \includegraphics[scale=.33, clip=true, trim = 0cm 1.25cm 0cm 0.61cm ]{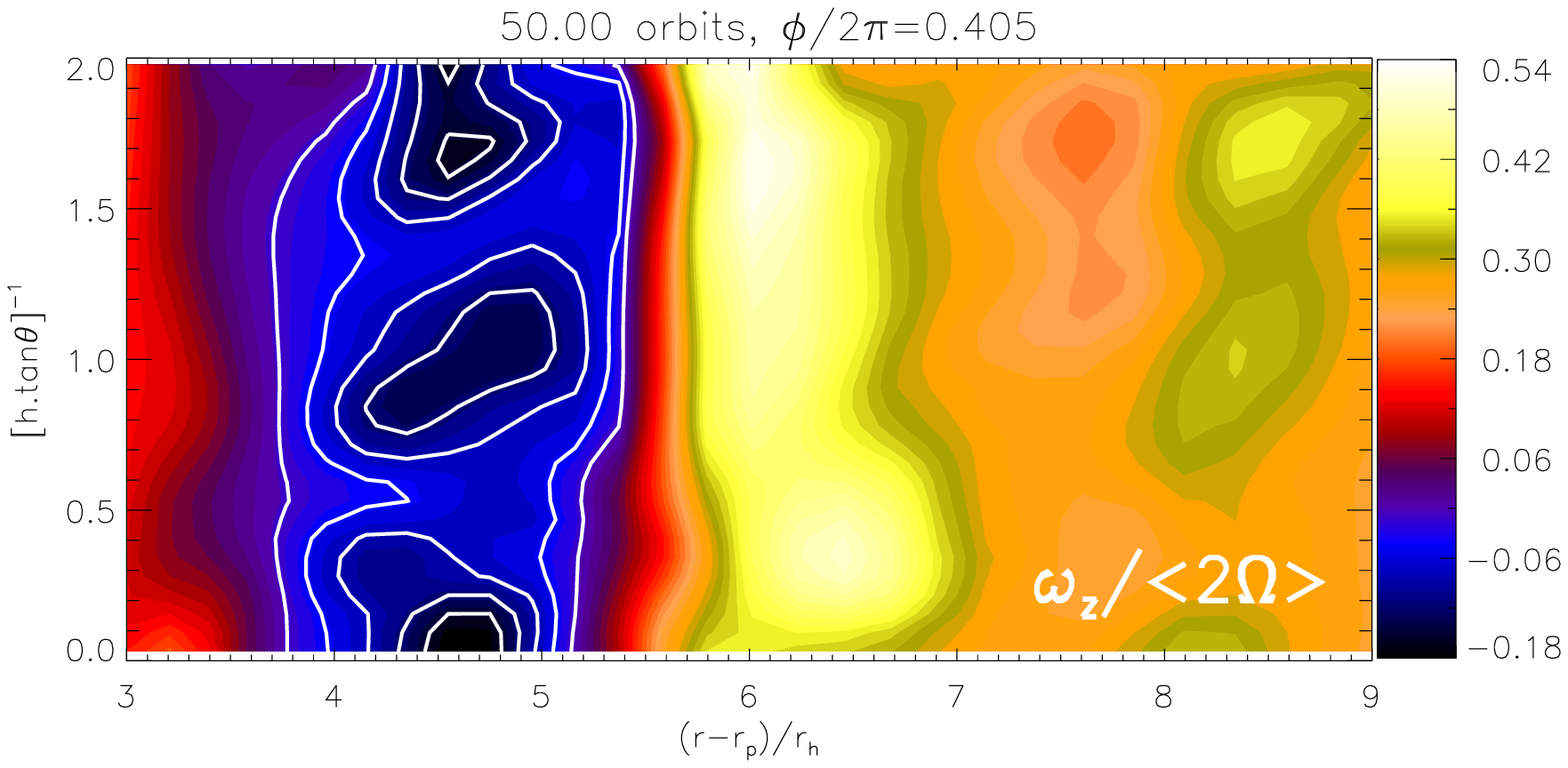}\\
    \includegraphics[scale=.33, clip=true, trim = 0cm 0.0cm 0cm 0.61cm ]{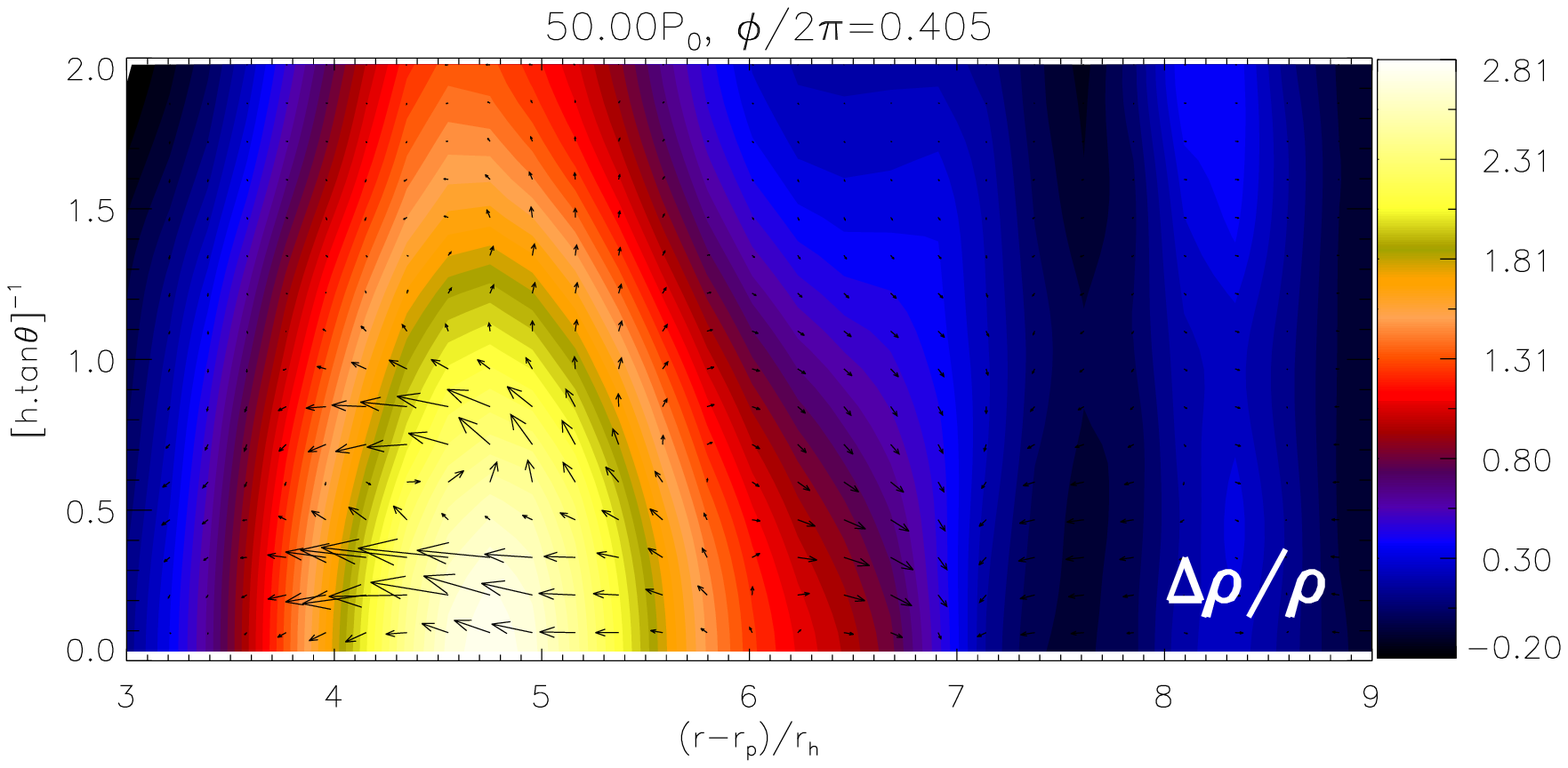}
  \end{minipage}
  \caption{Multi-vortex configuration at the end of a simulation for
    $Q_0=3$ ($t=50P_0$), with $h=0.07$ and $M_p=0.002M_*$. The Rossby
    number and  relative density perturbation in the $(r,\phi)$ plane
    at $\theta=\pi/2$ are shown on the left; and in the $(r,\theta)$
    plane at $\phi=\phi_0$ on the right, where $\phi_0$ is the vortex
    azimuth denoted by dotted lines.  
  }
  \label{vortex10}
\end{figure*}

Fig. \ref{vortex10} shows the relative density perturbation
and Rossby number at the end of the a simulation with
$Q_0=3$. Self-gravity is included. The 5-vortex configuration is
sustained from its initial development, unlike in the weakly
self-gravitating case where merging occurred over the same
time-scale. The preference for linear vortex modes with higher
azimuthal wavenumber $m$ with increasing strength of self-gravity was observed in
2D simulations \cite{lyra08,lin11a}, and persists in 3D. The smaller
vortices here are stronger than the single vortex in previous case,
with Rossby number $Ro\sim -0.2$ and the relative density perturbation
has significant vertical dependence.  

\subsubsection{Long term simulation}\label{q3b}
\begin{figure*}
  \centering
  \includegraphics[scale=.33,clip=true, trim = 0cm 0.0cm 0cm 0.92cm
  ]{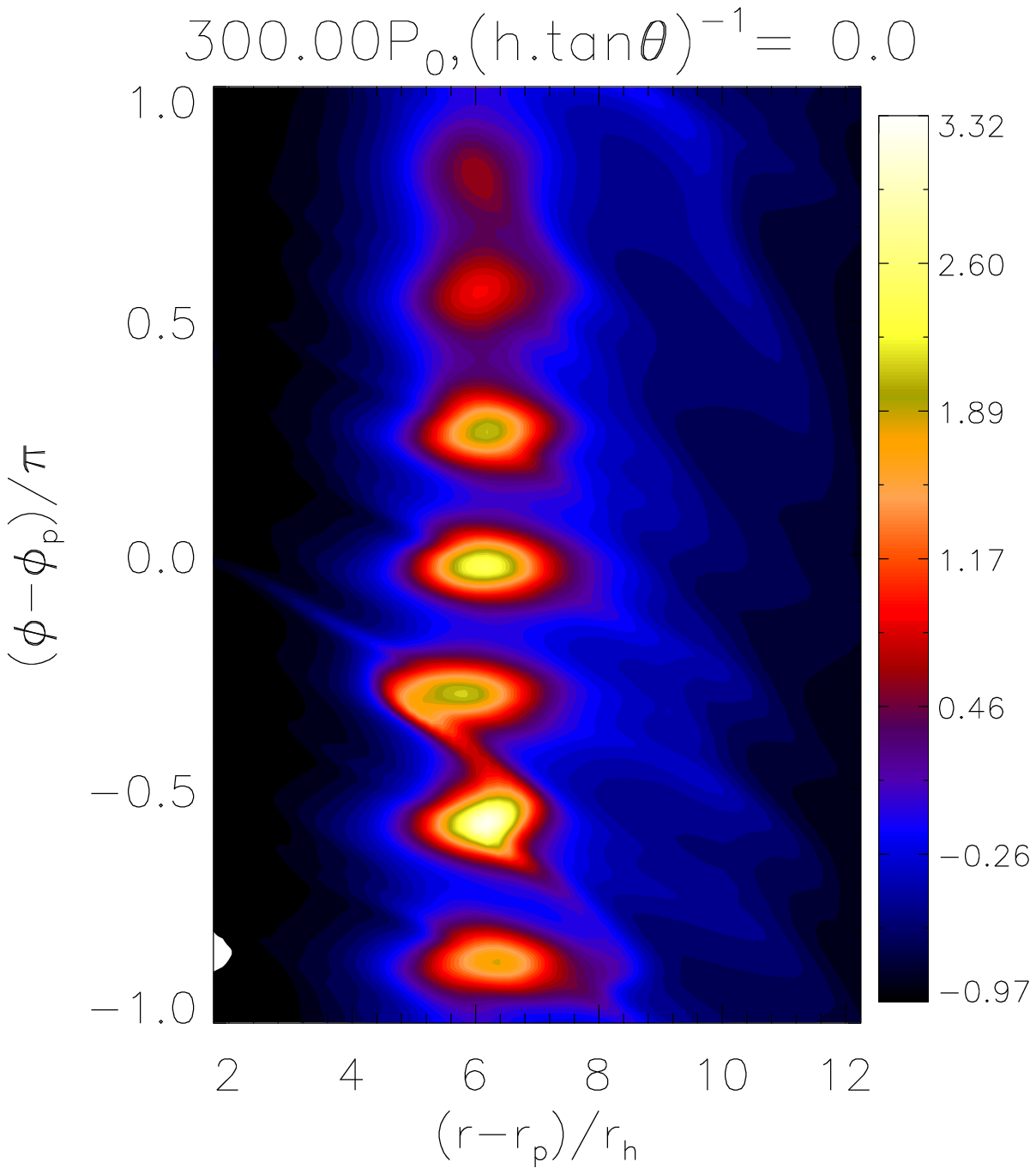}\includegraphics[scale=.33,clip=true, trim = 2.2cm 0.0cm 0cm 0.92cm
  ]{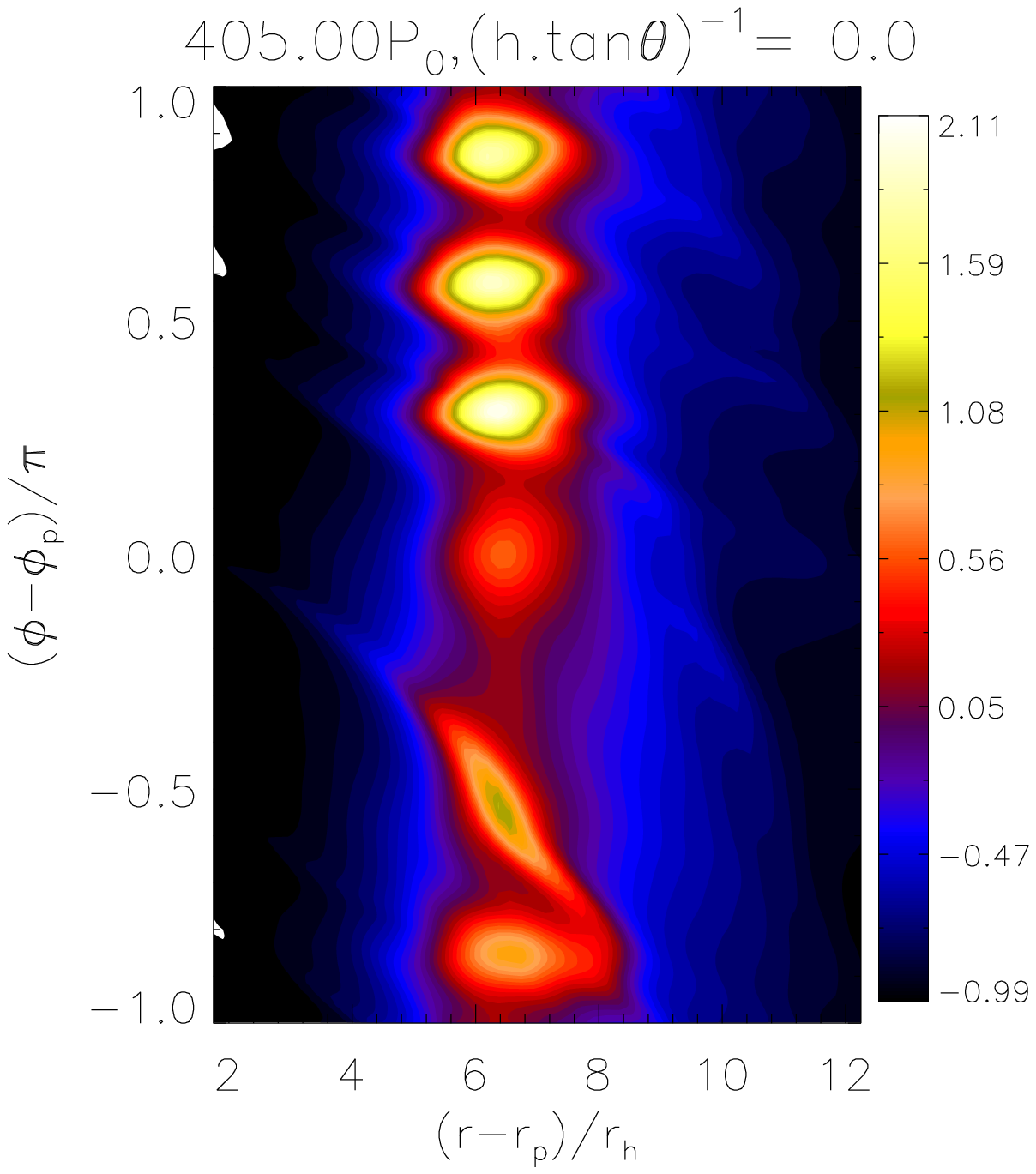}
  \includegraphics[scale=.33,clip=true, trim = 2.2cm 0.0cm 0cm 0.92cm
  ]{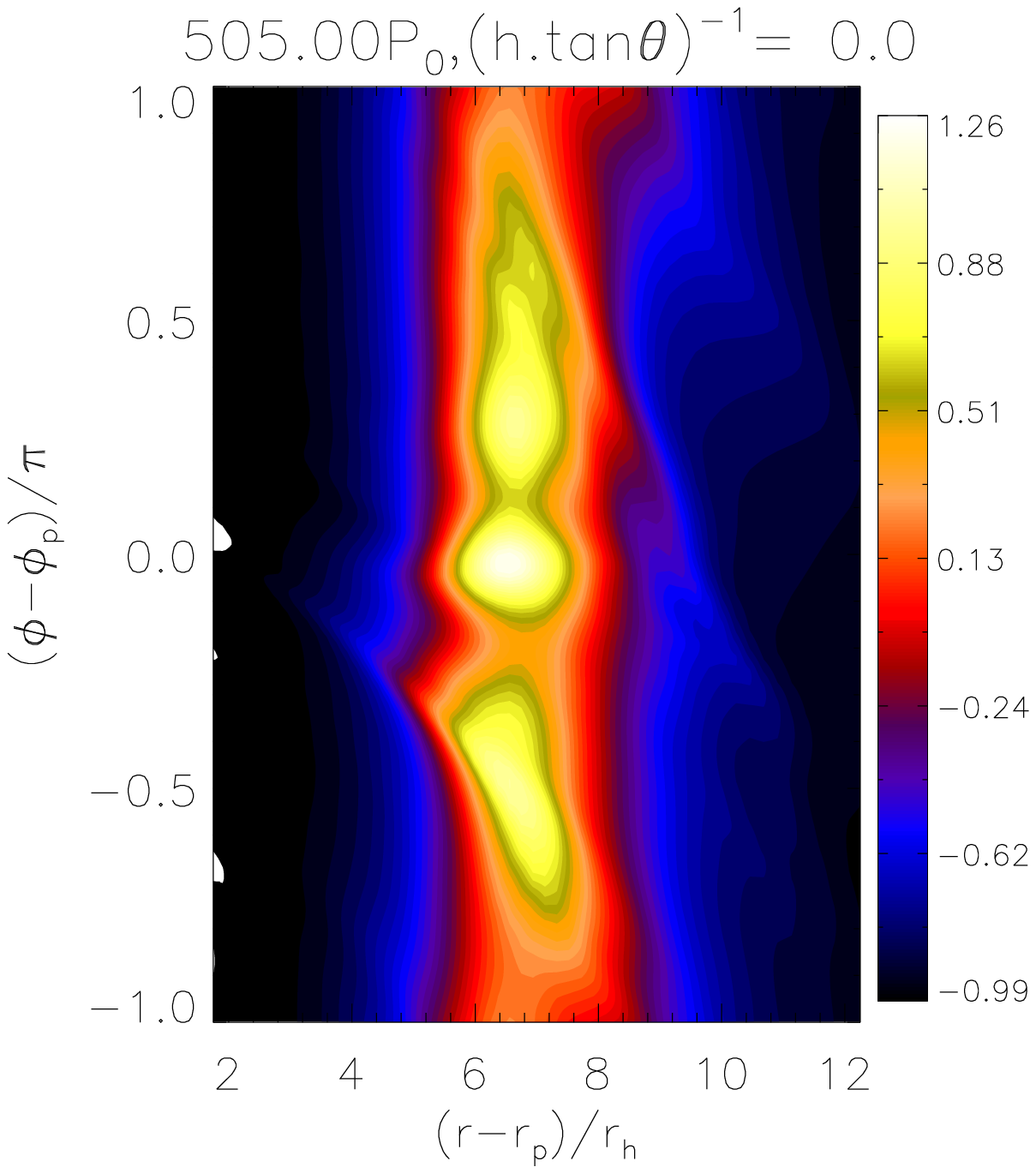}
  \caption{Simulation with $Q_0=3.0$, $h=0.05$ and $M_p=10^{-3}M_*$. The 
    midplane relative density perturbation is shown at $t=300P_0$
    (left), $t=405P_0$ (middle) and $t=505P_0$ (right). The minimum
    Rossby number was found to be $Ro=-0.11$ (left), $Ro=-0.09$
    (middle) and $Ro=+0.03$ (right).   
  }
  \label{vsg}
\end{figure*}

A smaller disk model, with $r\in[2,20]$, was simulated up to 
$t\sim500P_0$. $M_p=10^{-3}M_*$ and $h=0.05$ were adopted for this
run. Fig. \ref{vsg} shows the relative density perturbation towards the end
of the simulation. The multi-vortex configuration lasted $\sim 200$
orbits at the vortex radius. Notice a vortex may reach comparable
over-densities to the final post-merger vortex in the weakly
self-gravitating disk. It was observed that $|Ro|$  decreased from 0.2
at the onset of vortex formation, to $0.1$ towards the end of the
simulation, which may be due to limited numerical resolution.
 
\subsection{Gap edge spiral instability ($Q_0=1.5$)}\label{q1.5}
\begin{figure*}
  \centering
  \includegraphics[scale=.33,clip=true, trim = 1.24cm 1.0cm 2.72cm 0.94cm
  ]{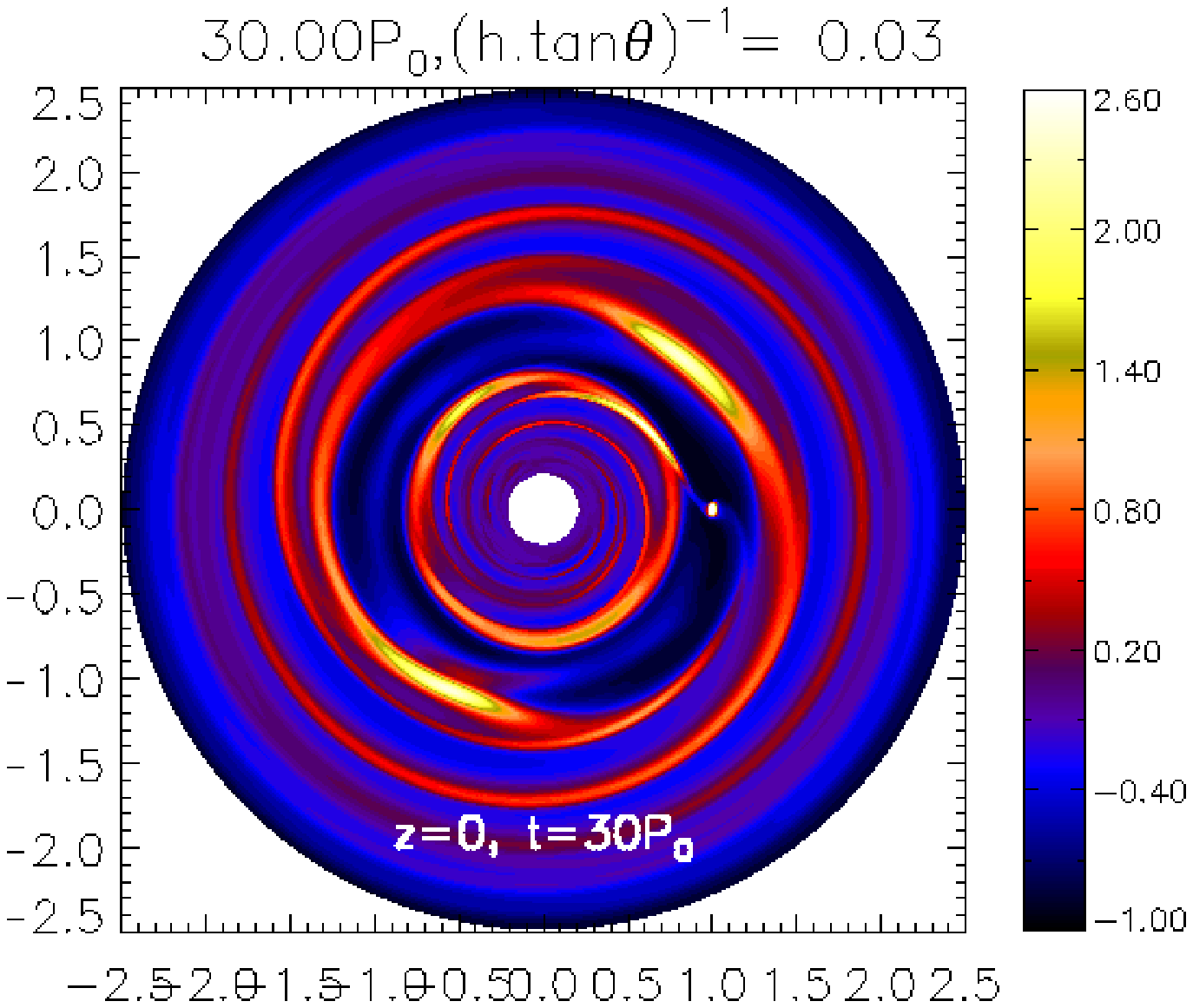}\includegraphics[scale=.33,clip=true, trim = 1.24cm 1.0cm 2.72cm 0.94cm
  ]{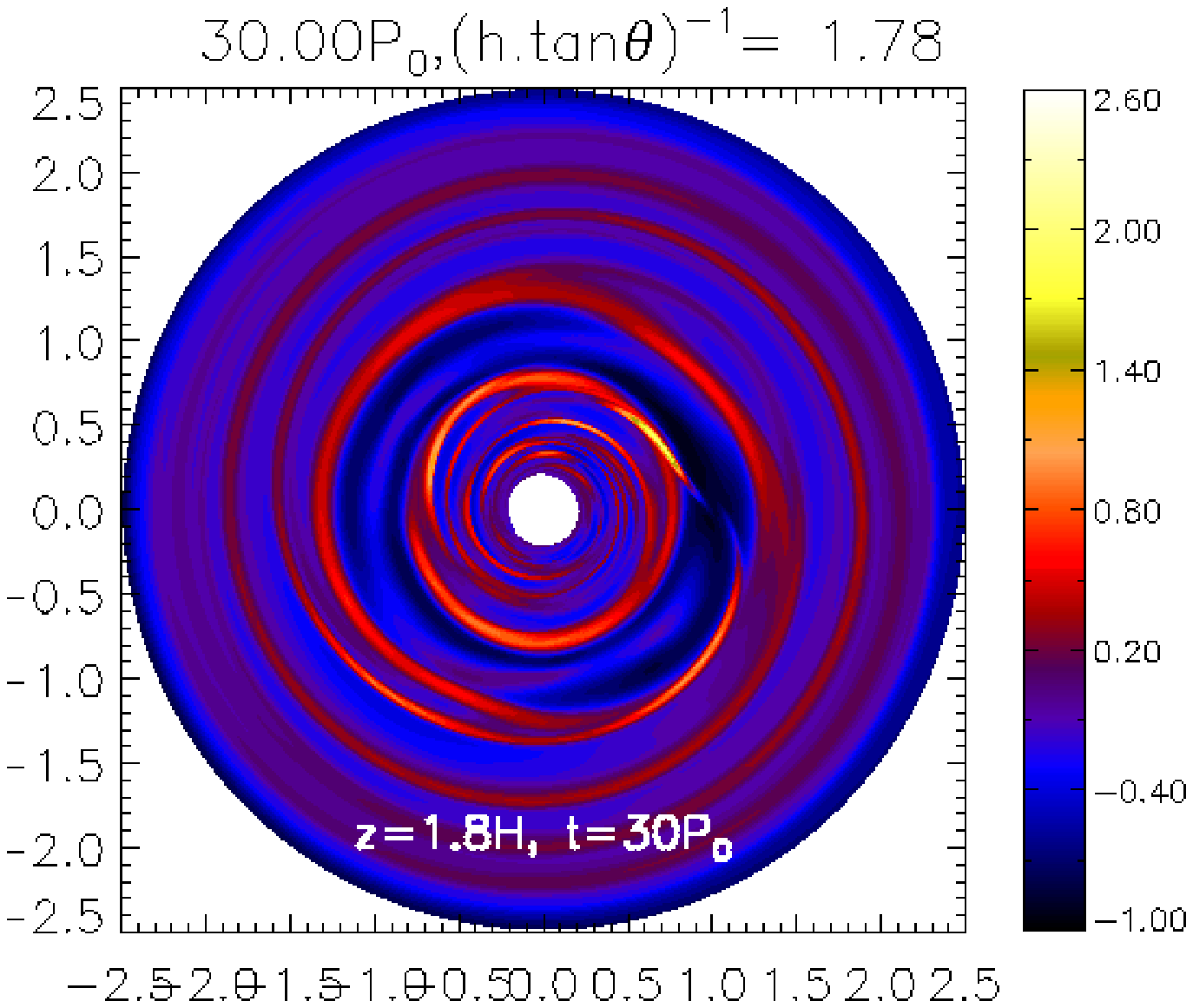}\includegraphics[scale=.33,clip=true, trim = 1.24cm 1.0cm 2.72cm 0.94cm
  ]{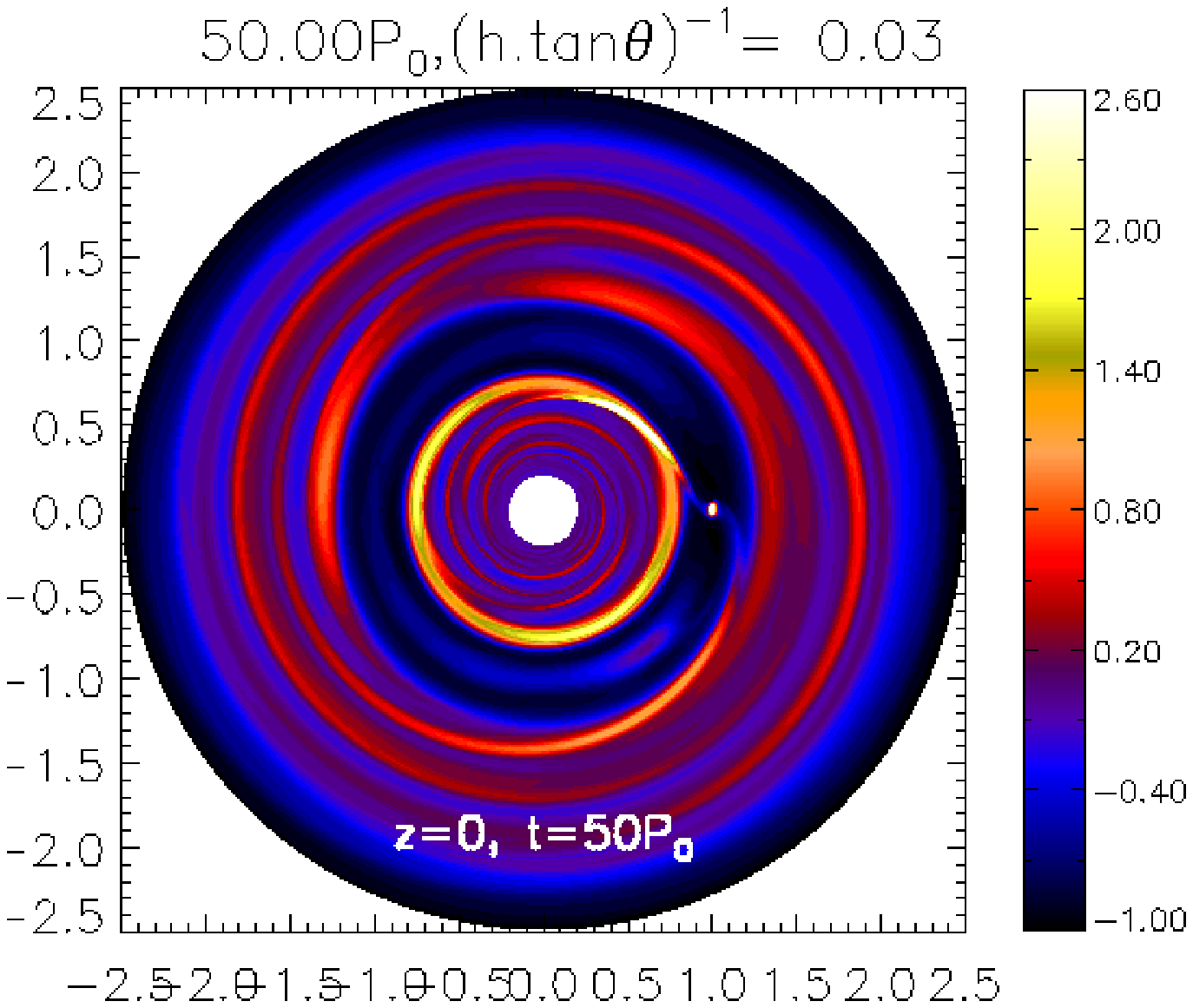}\includegraphics[scale=.33,clip=true, trim = 1.24cm 1.0cm 0cm 0.94cm
  ]{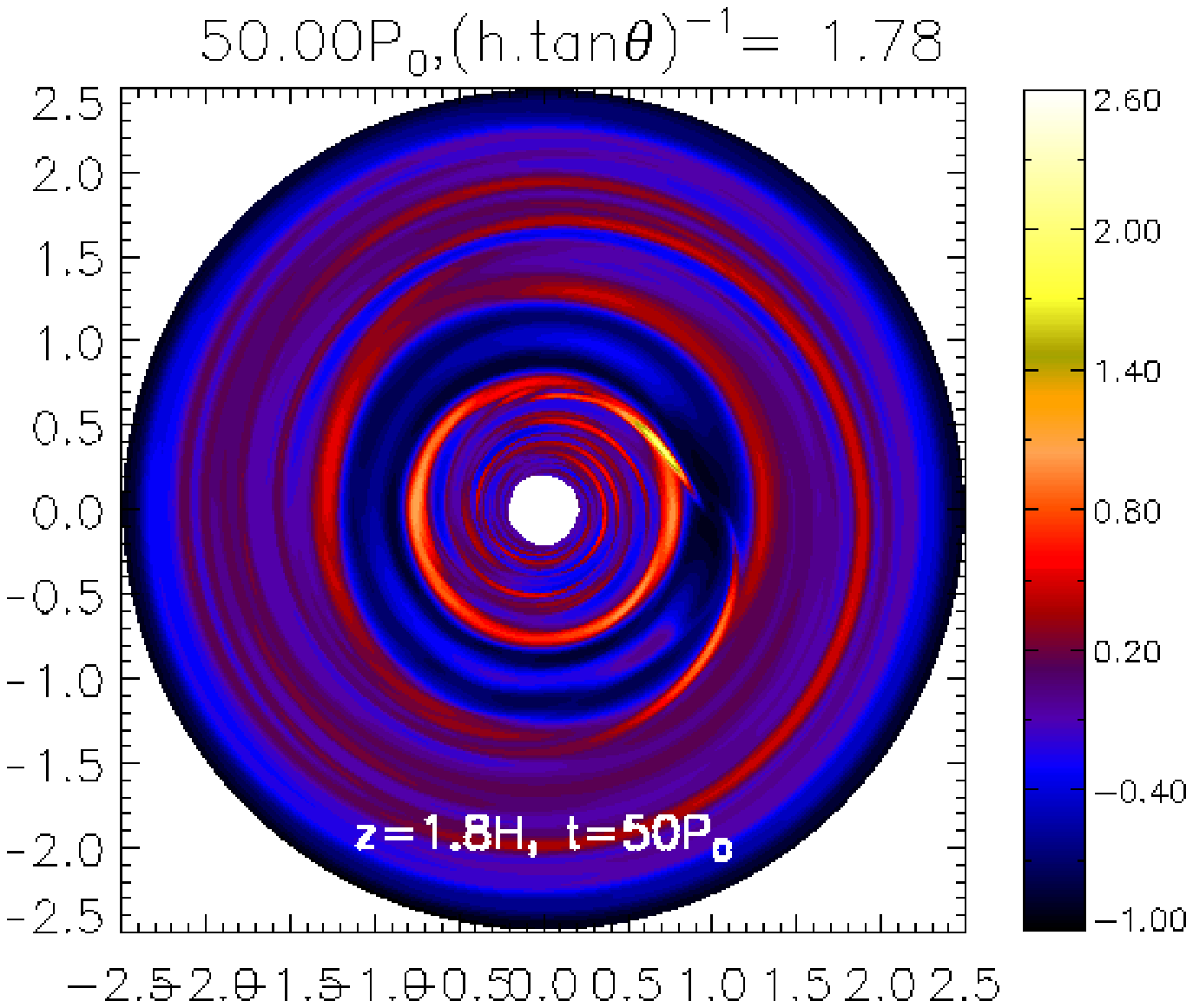}
  \caption{Spiral instability associated with the outer gap edge opened by a
    giant planet (\emph{not} a classic Toomre
    instability). The disk model is $Q_0=1.5$, $h=0.05$ with
    $M_p=10^{-3}M_*$. Self-gravity is sufficiently strong to suppress
    vortex formation. 
  }
  \label{vortex1}
\end{figure*}

The linear vortex instability can be suppressed by strong self-gravity. To
demonstrate this, a disk model with $Q_0=1.5$, $h=0.05$ and $M_p=10^{-3}M_*$
was simulated. Fig. \ref{vortex1} shows the development of an $m=2$
spiral mode associated with the outer gap edge. This instability
occurs during gap formation and supplies positive 
disk-on-planet co-orbital torques, because the over-density protrudes the outer gap edge and
approaches the planet from upstream. The disturbance is significantly
stratified, with most of the perturbation confined near the
midplane. The global spiral pattern appears transient, having 
decreased in amplitude by $t=50P_0$, but this is likely a radial
boundary condition effect.   

\section{Discussion}
Direct numerical simulations of 3D self-gravitating disk-planet
systems confirm the stability properties of gap edges previously
explored in 2D \citep{lyra08, meschiari08,
  lin11a,lin11b}. Vertical self-gravity enhances the  
density stratification of a vortex. Given the vortex instability
is only expected to occur in low viscosity regions of protoplanetary
disks --- dead zones --- which are overlaid by actively accreting
layers \citep{oishi09}, it may be advantageous to have self-gravity confining the
over-density near the midplane, thereby mitigate upper disk 
boundary effects and make the instability a more robust mechanism for
vortex formation.





\end{document}